\documentclass[aps,prc,twocolumn,floatfix,nofootinbib,superscriptaddress]{revtex4}
\usepackage{amsmath}
\usepackage{amsfonts}
\usepackage{amssymb}
\usepackage{bm}
\usepackage{graphicx}
\usepackage{multirow}
\usepackage{color}
\usepackage{mathrsfs}
\usepackage{bbold}
\usepackage{epsfig}
\usepackage{longtable}
\usepackage{nicefrac}

\usepackage{color}
\definecolor{purple}{rgb}{0.5,0,0.5}
\definecolor{blue}{rgb}{0.0,0,0.9}
\usepackage[colorlinks=true,linkcolor=blue,citecolor=blue,urlcolor=blue]{hyperref}

\usepackage{tabularx}
\newcolumntype{L}[1]{>{\raggedright\arraybackslash}p{#1}}
\newcolumntype{C}[1]{>{\centering\arraybackslash}p{#1}}
\newcolumntype{R}[1]{>{\raggedleft\arraybackslash}p{#1}}

\begin{document}

\title{Charge dependence of mesons with flavored contact-interaction couplings}

\author{F\'abio L. Braghin}
\email{braghin@ufg.br}
\affiliation{Instituto de F\'isica, Universidade Federal de Goias, Avenida Esperan\c{c}a, s/n, 74690-900, Goi\^ania, Goias, Brazil}

\author{Bruno El-Bennich}
\email{bennich@unifesp.br}
\affiliation{Departamento de F\'isica, Universidade Federal de S\~ao Paulo, Rua S\~ao Nicolau 210, Diadema, 09913-030 S\~ao Paulo, Brazil}
\affiliation{Instituto de F\'{\i}sica Te\'orica, Universidade Estadual Paulista, Rua Dr.\;Bento Teobaldo Ferraz 271, 01140-070 S\~ao Paulo, S\~ao Paulo, Brazil}

\author{Fernando E. Serna}
\email{fernando.serna@unisucrevirtual.edu.co}
\affiliation{Departamento de F\'isica, Universidad de Sucre, Carrera 28 No.~5-267, Barrio Puerta Roja, Sincelejo 700001, Colombia}

%%%%%%%%%%%%%%%%%%%%%%%%%%%%%%%%%%%%%%%%%%%%%%%%%%%%%%%%%%%%%%%%%%%%%%%%%%%%%%%%%%%%%%%
%%%%%%%%%%%%%%%%%%%%%%%%%%%%%%%%%%%%%%%%%%%%%%%%%%%%%%%%%%%%%%%%%%%%%%%%%%%%%%%%%%%%%%%

\begin{abstract}

Effective interaction models of quantum chromodynamics, based on quark degrees of freedom, have been successfully employed to compute the properties of a large array 
of ground and excited meson and baryon states, along with their electromagnetic form factors, distribution functions and thermal behavior. Amongst them, the contact-interaction 
model, while non-renormalizable, implements confinement, satisfies Lorentz covariance and correctly describes chiral symmetry and its dynamical breaking pattern. Original 
studies focused on the light hadron sector in the isospin limit and were thereafter extended to heavy mesons and baryons. The strong effective couplings, as well as infrared 
and ultraviolet regulators, are flavor-dependent model parameters adjusted to  reproduce hadronic observables. In contrast, in this study we combine SU(4) flavor-symmetry 
breaking couplings, obtained from one-loop vacuum polarization amplitudes in the presence of background constituent quark currents, with the contact-interaction model. 
This allows us to reduce the number of mass-dimensioned parameters and to consistently relate all flavored couplings to a single mass scale, while the masses and weak 
decay constants of the pions, kaons, $D$ and $D_s$ mesons are in good agreement with average reference values.  Allowing for realistic isospin breaking, $m_d/m_u = 1.7$, 
in conjunction with the effect of the flavored couplings, leads to a mass splitting, $m_{\pi^+}- m_{\pi^0} \approx 0.3$\;MeV, that agrees with lattice QCD values. For the kaons, 
the mass difference is $m_{K^0}- m_{K^\pm} \approx 2.3$\;MeV, whereas  $m_{D^\pm} - m_{D^0} \approx 0.5$\;MeV and the $\eta_c$ is 6\% lighter than the experimental mass.
\end{abstract}

\maketitle

%%%%%%%%%%%%%%%%%%%%%%%%%%%%%%%%%%%%%%%%%%%%%%%%%%%%%%%%%%%%%%%%%%%%%%%%%%%%%%%%%%%%%%%
%%%%%%%%%%%%%%%%%%%%%%%%%%%%%%%%%%%%%%%%%%%%%%%%%%%%%%%%%%%%%%%%%%%%%%%%%%%%%%%%%%%%%%%

\section{Introduction}
\label{sec:intro}

The current quark masses, and therefore the Higgs mechanism, are the unique source of flavor symmetry breaking (FSB) in quantum chromodynamics (QCD). 
In low-energy effective models of this theory, one should expect that all free model parameters manifest FSB in some form and therefore be flavor dependent.  
Flavor-dependent contact interactions induced by vacuum polarization were investigated in the Nambu--Jona-Lasinio~(NJL) model~\cite{Braghin:2021,Braghin:2021hmr,deSousa:2023ivb} 
considering two types of processes. The first process takes into account contributions from the one-loop quark determinant in the presence of background constituent quark  
currents and flavor SU(3)~\cite{Braghin:2021,Braghin:2021hmr}. In that approach, background quark currents are dressed by components of an effective gluon propagator  
which  yields constituent-quark currents. In a second approach, the one-loop vacuum polarization within the NJL model provides corrections to the coupling constant of 
the model~\cite{deSousa:2023ivb}.

The resulting quark-antiquark interactions in both approaches 
have been resolved in the long-wavelength limit and are written as flavor-dependent corrections to the NJL model coupling constants. These couplings can be calculated from 
a given set of effective quark masses that are, conversely, dependent on the quark-antiquark flavor-dependent interactions, leading to a self-consistent set of equations for effective 
masses and coupling constants. Small meson-mixing interactions due to FSB have only been partially studied in this context.

The quark-meson mixing that arises, due to different representations of the flavor group in which mesons and quarks are defined, is explicit in the flavor dependence of the coupling 
constants.  This additional flavor dependence in the NJL model leads to a sizable amplification of quark-mass differences and to a reduction of the effective quark masses. 
Their effects in light scalar quark-antiquark states have also been investigated without mixing interactions, and numerical results corroborate various studies pointing out that most 
scalar mesons cannot be exclusively  described by quark-antiquark states~\cite{Braghin:2022uih}.
  
In this work, the FSB couplings are introduced in a contact-interaction (CI) model that makes use of a covariant proper-time regularization and is characterized by a gluon-mass scale. 
The advantage of the CI model lies in the momentum independence of interaction kernels, avoiding integrals over momentum-dependent gluon propagators, which leads to simpler 
algebraic equations. At the same time, the model can be constructed to preserve chiral symmetry and its dynamical breaking, and translational invariance~\cite{Serna:2017nlr}. 
It therefore serves as a computationally efficient benchmarking tool for more complex, realistic QCD-based models, especially when considering a large set of observables, to 
understand the qualitative structure of hadrons. 

We aim at making a more consistent use of the CI model by reducing the number of free mass-dimensioned parameters in computing the mass spectrum of light and heavy mesons. 
To that end, the aforementioned flavored couplings are expressed in terms of a common flavor-independent mass parameter and inserted in the Dyson-Schwinger equation (DSE) of 
the quark and in the quark-antiquark interaction kernel of the Bethe-Salpeter equation (BSE). Conversely, we implement a common proper-time regularization that relies on merely 
one set of infrared and ultraviolet regulators for all flavors considered. Despite these restrictions, we obtain masses and weak decay constants of the pions, kaons and $D$ mesons 
that depend on their charge state and are in good agreement with experimental values. The heavy quarkonium $\eta_c$, however, poses a challenge in this scheme.

The remainder of this paper summarizes the application of the CI model in the quark gap equation and the bound-state equation for pseudoscalar mesons in Section~\ref{PSmesonCI}, 
while the flavor-dependent couplings are introduced and their calculation is schematically described in Section~\ref{FSBcoupling}. We present our numerical results for the masses 
and weak decay constants of the $\pi^\pm$, $\pi^0$, $K^\pm$, $K^0$, $D^\pm$, $D^0$, $D_s$ and $\eta_c$ in Section~\ref{results}, and we close with a summary and a few 
comments about future perspectives in Section~\ref{lastsec}.

%%%%%%%%%%%%%%%%%%%%%%%%%%%%%%%%%%%%%%%%%%%%%%%%%%%%%%%%%%%%%%%%%%%%%%%%%%%%%%%%%%%%%%%

\section{Pseudoscalar mesons in the contact-interaction model\label{PSmesonCI}}

%%%%%%%%%%%%%%%%%%%%%%%%%%%%%%%%%%%%%%%%%%%%%%%%%%%%%%%%%%%%%%%%%%%%%%%%%%%%%%%%%%%%%%%

\subsection{Gap Equation}
\label{gapeq}

The CI model we consider is discussed in Ref.~\cite{Hellstern:1997nv,Gutierrez-Guerrero:2010waf}, for instance, and implements a confining,
symmetry-preserving treatment of a vector-vector CI.  In essence, the model is related to the NJL model as the quark-quark interactions are point-like, 
though by incorporating the nature of a vector-boson-exchange theory they are of the vector-vector type. Practically, this is implemented by reducing 
the finite-range gluon exchange between light quarks to a CI defined as, 
\begin{equation}
  g^2 D_{\mu \nu}(p-q)=  \frac{\delta_{\mu \nu}}{m_g^2} \, ,
\label{CIdef}  
\end{equation}
where $m_g \sim 0.5-1.5$\;GeV is a gluon-mass scale that is generated dynamically in QCD~\cite{Cornwall:1981zr,Aguilar:2004sw,Aguilar:2008xm,Aguilar:2011yb,Aguilar:2022thg}.

In a functional approach to QCD, the fully dressed quark propagator $S_f (p)$ of flavor $f$ in Euclidean space is the solution of the gap equation, 
\begin{align}
    S^{-1}_f  (p)  &= \  Z_2\, i\gamma\cdot p + Z_4 m_f  \nonumber \\
   &  +   \ Z_1 g^2 \!\! \int_q \, D_{\mu\nu}^{ab} (q-p) \,\frac{\lambda^a}{2}\gamma_\mu\,  S_f(q) \, \Gamma^{b}_\nu (q,p)\, , 
\label{eq:DSEqp}
\end{align}
where $\int_q = \int^\Lambda d^4 q /(2 \pi)^4$ is a mnemonic shortcut,  $m_f$ is the renormalized Lagrangian current-quark mass and $Z_1(\mu,\Lambda)$ and 
$Z_2(\mu,\Lambda)$ are the vertex and wave-function renormalization constants at the scale $\mu$. In the self-energy term, $D_{\mu\nu}$ is the dressed-gluon  
propagator and $\Gamma^a_\mu (k,p) = \frac{1}{2}\,\lambda^a  \Gamma_\mu (k,p)$ is the quark-gluon vertex~\cite{Serna:2018dwk,Albino:2021rvj,
El-Bennich:2022obe,Lessa:2022wqc,El-Bennich:2024hmy}, where $\lambda^a$ are SU(3) color matrices. 

The general covariant solution of the gap equation is,
\begin{equation}
    S _f (p)  =  \frac{ Z_f (p^2 )}{ i \gamma \cdot p + M_f ( p^2 )}\, ,
\label{DEsol}                        
\end{equation}
in which $Z_f (p^2)$ is the  wave renormalization function and $M_f (p^2)$ the mass function of the quark, respectively.
Employing the CI as defined in Eq.~\eqref{CIdef}, limiting the quark-gluon vertex to its leading truncation $\Gamma_\mu (k,p) = \gamma_\mu$,  
and taking the color trace, the DSE of Eq.~\eqref{eq:DSEqp} simplifies:
\begin{equation}
   S^{-1}_f (p)  = i \gamma \cdot p+m_f +\frac{4}{3} \frac{1}{m_{g,f}^2} \int_q \gamma_\mu S_f(q) \gamma_\mu \, .
 \label{CI-DSE}  
\end{equation}
We omit the renormalization constants, as the integral now possesses a quadratic divergence and the regularization procedure must therefore be adapted
as will be seen shortly. Note that we allow for a flavor dependence of the effective ``gluon mass'' $m_{g,f}$, as discussed in Ref.~\cite{Serna:2017nlr},
which is important to account for the fact that heavy-flavor quarks probe shorter distances than light-flavor quarks at the corresponding quark-gluon vertices. 
This flavor dependence has been successfully employed in the past years in the context of CI models~\cite{Gutierrez-Guerrero:2019uwa,Gutierrez-Guerrero:2021rsx,
Sultan:2024hep} and in finite-interaction DSE  calculations~\cite{El-Bennich:2016qmb,Chen:2019otg,Qin:2019oar,Serna:2020txe,Serna:2022yfp,daSilveira:2022pte,Serna:2024vpn}.

The general solution in Eq.~\eqref{DEsol} is modified to a constant-mass propagator, 
\begin{equation}
   S_f(p) = \frac{1}{i \gamma \cdot p + M_f } \, ,
 \label{CI-propagator}  
\end{equation}
which is a consequence of the lack of relative momentum transfer in the interaction. The momentum-independent constituent mass $M_f$ is obtained from the integral, 
\begin{equation}
   M_f =  m_f +  \frac{16}{3} \frac{M_f}{m_{g,f}^2}  \int_q\, \frac{1}{q^2+M_f^2 } \, .
  \label{massfuncCI} 
\end{equation}
It is noteworthy that the values of $M_f$ are in agreement with typical constituent quark masses~\cite{ElBennich:2008xy,ElBennich:2008qa,daSilva:2012gf,
ElBennich:2012ij,deMelo:2014gea,El-Bennich:2009gbu,El-Bennich:2008dhc}. However, $M_f$ is dynamically generated in the CI model. 

At this point, we must regularize the quadratic divergences, and a common procedure~\cite{Ebert:1996vx} is to rewrite the denominator as an integral, 
\begin{equation}
\hspace*{-2mm}
   \frac{1}{q^2+M_f^2}  =\int_0^{\infty} \!\! d \tau\, \mathrm{e}^{-\tau (q^2 +M_f^2 )} \Rightarrow \int_{\tau_{\mathrm{uv}}^2}^{\tau_{\mathrm{ir}}^2} \!\! d \tau \,
   \mathrm{e}^{-\tau (q^2 +M_f^2 )} ,
 \label{properprescript}  
\end{equation}
which introduces infrared and ultraviolet regulators, $\tau_{\mathrm{ir}}$ and $\tau_{\mathrm{uv}}$, respectively. The energy scale 
$\Lambda_{\mathrm{uv}}^2=1 / \tau_{\mathrm{uv}}$ is a model parameter that sets the scale of a given physical system, while a finite value of  
$\Lambda_{\mathrm{ir}}^2 =  1/\tau_{\mathrm{ir}}$ implements confinement by avoiding quark thresholds. With this regularization scheme, we can 
write Eq.~\eqref{massfuncCI} as,
\begin{equation}
  M_f =m_f +   \frac{M_f}{3 \pi^2  m_{g,f}^2} \,  \mathcal{F}\left(M_f^2 , \tau_{\mathrm{ir}}, \tau_{\mathrm{uv}}\right) \, ,
\end{equation}
where $\mathcal{F}(M_f^2 , \tau_{\mathrm{ir}}, \tau_{\mathrm{uv}})$ is defined as the difference of incomplete gamma functions $\Gamma(\alpha, y)$:
\begin{equation}
  \frac{ \mathcal{F}(M_f^2 , \tau_{\mathrm{ir}}, \tau_{\mathrm{uv}} ) }{ M_f^2} \equiv \Gamma (-1, M^2_f \tau_{\text {uv }}^2 )-\Gamma (-1, M^2_f \tau_{\text {ir }}^2 ) \, . 
\end{equation}
Finally, in both Eq.~\eqref{CI-DSE} and Eq.~\eqref{eq:BSEps}, we employ the \emph{same} ultraviolet cutoff $\Lambda \equiv \Lambda_{\mathrm{uv}}$.

%%%%%%%%%%%%%%%%%%%%%%%%%%%%%%%%%%%%%%%%%%%%%%%%%%%%%%%%%%%%%%%%%%%%%%%%%%%%%%%%%%%%%%%

\subsection{Bethe-Salpeter Equation}

The mass spectrum and wave functions of the pseudoscalar mesons are obtained by solving an eigenvalue problem posed by the homogeneous BSE,
\begin{equation}
\hspace*{-1mm}
 \Gamma^{ef}_M (k,P)  = \! \int_q K_{ef}(k,q,P)  S_f (q_+)\Gamma^{ef}_M(q,P)  S_g (q_-) \,  ,
\label{eq:BSEps}
\end{equation}
The flavor indices, $ef$, denote the quark content of the meson, while $q_+ = q + \alpha P$,  
$q_{-} = q - (1-\alpha) P$ are the quark  momenta and $0 \leq \alpha \leq 1$ is an arbitrary momentum-partition parameter.
We follow Refs.~\cite{Serna:2020txe,Qin:2019oar,Serna:2017nlr} in using an \emph{average} of interactions, defined for each flavor with 
the DSE~\eqref{CI-DSE}, in the BSE kernel,
\begin{equation}
 K_{ef} =  - \frac{1}{m_{g,f} m_{g,e} }\, \frac{\lambda^a}{2} \gamma_\mu   \frac{\lambda^a}{2} \gamma_\mu \, ,
\label{eq:Kcontact}
\end{equation}
which is momentum independent. The Bethe-Salpeter amplitude (BSA) is the sum of two covariant amplitudes, equally independent of the relative momentum $k$, 
\begin{equation}
  \Gamma^{ef}_M (P)=\gamma_5\left [ \, i E^{ef}_M(P) + \frac{1}{2M_{ef}}\,  \gamma \cdot P\,F^{ef}_M(P) \right ] ,
\label{contactBSA}
\end{equation}
where $M_{ef}=M_e M_f/(M_e+M_f)$. Therefore, the  BSE can be represented as a $2 \times 2$ matrix $\mathcal{K}$ of the form,
\begin{equation}
\hspace*{-0.3cm}
\left [ \!
\begin{array}{c}
  E^{ef}_M(P)\\[0.2true cm]
  F^{ef}_M(P)
\end{array}
\! \right ]
  =   \frac{1}{3m_{g,e} m_{g,f} } 
\left [ 
\begin{array}{cc}
\!
  \mathcal{ K}^{EE}_M & \mathcal{ K}^{EF}_M \\[0.2true cm]
  \mathcal{K}^{FE}_M & \mathcal{ K}^{FF}_M
\end{array} 
\! \right ]
\! 
\left [ \!
\begin{array}{c}
  E^{fg}_M(P)\\[0.2true cm]
  F^{fg}_M(P)
\end{array}
\! \right ] \! ,
\end{equation}
and the matrix elements of the kernel are given by~\cite{Serna:2017nlr},
\begin{eqnarray}
\mathcal{ K}^{EE}_M &=& - \int_q  \operatorname{Tr} \left [ \gamma_5  \gamma_\mu S_e (q_+)  \gamma_5 S_f (q_-) \gamma_\mu \right ],
\label{EE}  \\ %[0.2true cm]   
\mathcal{ K}^{EF}_M &=& \frac{i}{2M_{ef}} \int_q  \operatorname{Tr} \left [ \gamma_5 \gamma_\mu \, S_e (q_+) \gamma_5  {\gamma\cdot\! P} \,
S_f (q_-)  \gamma_\mu \right ], 
\label{EF} \\ %[0.2true cm] 
\mathcal{ K}^{FE}_M &=& \frac{2iM_{ef}}{P^2}\int_q  \operatorname{Tr} \left [ \gamma_5  {\gamma\cdot P}  \gamma_\mu S_e (q_+) \gamma_5 
S_f (q_-)  \gamma_\mu \right ] ,
\label{FE} \\ %[0.2true cm] 
\mathcal{ K}^{FF}_M &= & \frac{1}{P^2}\int_q  \operatorname{Tr}\left [ \gamma_5 {\gamma\cdot\! P} \gamma_\mu  S_e (q_+)\gamma_5 
 {\gamma\cdot P}  S_f (q_-) \gamma_\mu \right ] .  \hspace{7mm}
\label{FF} 
\end{eqnarray}
In Eqs.~\eqref{EE}--\eqref{FF} the traces are over Dirac indices and all integrals give rise to ultraviolet logarithmic and quadratic 
divergences~\cite{Gutierrez-Guerrero:2010waf,Serna:2017nlr}. 

The eigenvalue problem can thus be expressed as: 
\begin{equation}
 \lambda   ( P^2)\, \Gamma_M  (P^2 ) =  \mathcal{K} (P^2)\, \Gamma_M  (P^2) \, ,
\end{equation}
The BSAs are normalized canonically with the integral, 
\begin{align}
  2 P_\mu & =  N_c \! \int_q
  \operatorname{Tr}_D  \left [ \bar \Gamma_M^{ef}(-P) \frac{\partial S_e (q_{+} )}{\partial P_\mu} \, \Gamma_M^{ef}(P) S_f (q_{-} ) \right.  \nonumber \\
 &    \left.  + \ \Gamma_M^{ef}(-P) S_e (q_{+} ) \Gamma_M^{ef} (P) \frac{\partial S_f (q_{-} ) }{\partial P_\mu}  \right ] ,
\end{align}
where $\bar \Gamma_M (P) = C \Gamma^T_M (P) C^T $ is the charge-conjugated BSA.  We remind that the vast majority  of NJL models merely retain the 
$E^{ef}_M(P) $ component of the BSA, otherwise known as the random-phase approximation of the BSE~\cite{Vogl:1991qt,Klevansky:1992qe,Bijnens:1995ww}. 
The proper normalization of the BSA allows to compute the weak decay constant of a pseudoscalar meson:
\begin{equation}
\hspace*{-1.5mm}
   f_M\, p_\mu  =  \frac{Z_2 N_c}{\sqrt{2}} \! \int_q \operatorname{Tr}_\text{\tiny D} \big [\gamma_5 \gamma_\mu S_e (q_+)
                          \Gamma_M^{ef} (q,p)  S_f (q_-) \big ] ,
\label{pion_cte_DS}                      
\end{equation}
where the above trace is over Dirac indices and $N_c$ is the number of colors. The validity of the BSE kernel~\eqref{eq:Kcontact} is verified 
with the Gell-Mann-Oakes-Renner (GMOR) relation, which expresses the axial vector Ward-Green-Takahashi identity that describes axial vector current 
conservation. The leptonic decay constants of the light mesons and of the $D$ mesons obtained with the GMOR relation agree within less than 0.01\% 
with those listed in Tab.~\ref{table1}.

%%%%%%%%%%%%%%%%%%%%%%%%%%%%%%%%%%%%%%%%%%%%%%%%%%%%%%%%%%%%%%%%%%%%%%%%%%%%%%%%%%%%%%%
%%%%%%%%%%%%%%%%%%%%%%%%%%%%%%%%%%%%%%%%%%%%%%%%%%%%%%%%%%%%%%%%%%%%%%%%%%%%%%%%%%%%%%%

 \section{Flavor-dependent couplings\label{FSBcoupling}}

In Section~\ref{gapeq} we introduced the FSB couplings in the CI model, which are crucial to describe heavy-light  mesons that contain two very distinct 
mass scales. This is due to the fact that heavy-flavor quarks probe shorter distances than light-flavor quarks at the corresponding quark-gluon vertices. 
Therefore, one  expects a  smaller coupling strength for heavy-flavor quarks.

One may define couplings for the flavor sectors in the gap equation, for instance the coupling $g_i \equiv 1/m_{g,i}^2$ for the light quarks, $i = u,d,s$, 
and another coupling for the charm sector, $g_c \equiv 1/m_{g,c}^2$, and this is indeed how the CI model is mostly applied to spectroscopy or form factor 
calculations. The couplings along with the flavor-dependent cutoffs $\Lambda_{\mathrm{ir}}$ and $\Lambda_{\mathrm{uv}}$ and quark masses $m_u = m_d$, 
$m_s$ and $m_c$  are the model parameters, adjusted to reproduce the masses and leptonic decay constants of  the pion, kaon, $D$ and $D_s$ mesons and 
$\eta_c$. At this level,  the CI model is not able to distinguish between charge states of the mesons.

The aim of the present study is to relate the flavored couplings to a single mass scale $g_0 \equiv 1/m_g^2$ and a unique set of ultraviolet and infrared cutoffs 
$\Lambda_{\mathrm{uv}}$ and $\Lambda_{\mathrm{ir}}$, respectively. To that end, we employ interactions induced by vacuum polarization processes which were
proposed in Refs~\cite{Braghin:2021,Braghin:2021hmr}. They can be obtained by considering the one-loop background field method as shortly described below. 
Beforehand, it is worthwhile to remark that the background-field mechanism should be associated with a two-gluon exchange process rather than with a flavored 
one-gluon effective mass interaction. More precisely, these interactions have been shown to correspond to a two-gluon exchange obtained from a one-loop quark 
determinant in the presence of background constituent-quark currents. The long-wavelength limit of these interactions can then be used as corrections to punctual 
quark-antiquark effective interactions. The exact two-gluon exchange calculation has not explicitly been carried out yet, though it also leads to flavor-dependent interactions.

Consider the Lagrangian that defines the CI,
\begin{equation} 
    \mathcal {L}_I = \frac{g_0}{2} \left (\bar{\psi} \lambda_i \gamma_\mu \psi \right )\!  \left  (\bar{\psi} \lambda_i \gamma^\mu \psi \right )   , 
\label{vector-interaction}
\end{equation}
where $\lambda_i $ are the SU(4) flavor matrices in the adjoint representation. Employing the one-loop background field method with quantum and classical background 
quark currents, denoted by the quark fields $\psi_1$ and $\psi_2$ respectively, Eq.~\eqref{vector-interaction} can  be cast in the form:
\begin{align} 
  \mathcal{L}_I  = \frac{g_0}{2} \left[ (\bar{\psi} \lambda_i \gamma_\mu \psi )_2   \right. &  (\bar{\psi} \lambda_i \gamma^\mu \psi )_2 
               +   (\bar{\psi} \lambda_i \gamma_\mu \psi )_1  (\bar{\psi} \lambda_i \gamma^\mu \psi )_1  \nonumber \\
               + &  \left. 2 (\bar{\psi} \lambda_i \gamma_\mu \psi )_1  (\bar{\psi} \lambda_i \gamma^\mu \psi )_2 \right ] .
\label{BFM-L}
\end{align}
The interactions of the background fields $\psi_2$ represent the original interactions to which one adds the polarization process, whereas
the interaction of quantum fields $\psi_1$ can introduce local quark-antiquark meson fields, which are neglected at one-loop level.  
The effective quark determinant for the remaining terms is then obtained as,
\begin{align}  \hspace*{-2mm}
    S_{\text{eff}}   = C_0   + \tfrac{i}{2}  \operatorname{Tr}_{\text{CDF}}  & \, \ln
  \left\{ \! \big ( 1 +  S_f(k)  g_0  \lambda_i \gamma_\mu   (\bar{\psi} \lambda_i \gamma_\mu \psi )_2  \big )^\dagger \right.   \nonumber  \\
%\
  \times \,  \big ( 1 + & \! \left. S_f(k) g_0  \lambda_i \gamma_\mu   (\bar{\psi} \lambda_i \gamma_\mu \psi)_2  \big )  \right  \}  ,
\label{exp-1}
\end{align}
where the trace is over Dirac, color and flavor indices, the quark propagator was defined in Eq.~\eqref{CI-propagator} and $C_0$ is an irrelevant constant term. 
The index $_2$ is neglected from here on. 

Since the constituent-quark mass is large, we expand this determinant up to fourth order in the quark fields and determine the effective couplings with the 
zero order derivative expansion~\cite{Braghin:2021hmr,deSousa:2023ivb,Braghin:2024lbo,Braghin:2024yql}. To fourth order, the following effective Lagrangian 
is derived in the local limit of zero-momentum transfer,
\begin{equation}
   \mathcal{L}_{\text{eff}}^{(4)}   =   g_{ij} ( \bar \psi \lambda_i \gamma_\mu \psi)  (\bar \psi  \lambda_j \gamma_\mu \psi) +\,   \ldots\, ,
\end{equation}
where the dots stand for higher order derivatives and non-derivative quark-current terms and the flavor-dependent couplings in the $t$-channel are given by 
the polarization integrals, 
\begin{equation}
  g_{ij}  = g_0^2 \frac{i N_c}{8}\, \delta_{\mu\nu} \operatorname{Tr}_{\text{DF}}  \int_k  (S_f (-k)  \lambda_i \gamma_\mu  )^{\dagger}  S_e (k) \lambda_j \gamma_\nu \, .
 \label{flavorcoupl}
\end{equation}
The latter reduces to a set of integrals  
of the following type in Euclidean space:
\begin{equation}
   I_{ef} = N_c  g_0^2
   \int_k \frac{ k^2 + M_e M_f }{ (k^2 + M_e^2 ) ( k^2 + M_f^2 ) } \, .
 \label{integralcouple}  
\end{equation}
We consistently use the same proper-time regularization~\eqref{properprescript} of the propagators as in the DSE~\eqref{CI-DSE} and BSE~\eqref{eq:BSEps}.

We can now write an effective interaction Lagrangian that accounts for corrections due to the flavor-dependent couplings of the charged 
$\bar \psi \lambda_i \gamma_\mu \psi$  states:
\begin{equation}
   \mathcal{L}_{\text{eff}}=  \frac{ g_0 + g_{ij}}{2} (\bar{\psi} \lambda_i \gamma_\mu \psi )  (\bar{\psi} \lambda_j \gamma^\mu \psi ) \, , 
 \label{efflagrangianU4}  
\end{equation}
with $i,j = 0,1, \ldots, 8$. We note that whereas the diagonal couplings $g_{ii}$ receive corrections for all flavor states $i=1,\ldots, N_f^2$, nondiagonal couplings only 
exist between the diagonal currents $i,j={0,3,8}$, i.e. $g_{03}, g_{08}$ and $g_{38}$. These interactions are responsible for meson mixings, e.g. $\pi^0-\eta-\eta'$, 
as discussed  in Ref.~\cite{Braghin:2021,Braghin:2021hmr}, similarly to tree level calculations that include the 't Hooft determinant in the NJL model.

It follows from the invariance under U(1) and CP symmetry transformations that, 
\begin{align}
   g_{11} & = g_{22} \, , \ \  g_{44}=g_{55}\, , \ \  g_{66} = g_{77} \, ,  \nonumber \\ 
  g_{99} & = g_{1010} \, , \ \  g_{1111}  = g_{1212}\, ,  \ \  g_{1313} = g_{1414} \, ,
\end{align}
which ensures that charged mesons and antimesons have equal properties. In terms of the integrals in Eq.~\eqref{integralcouple}, the flavored 
couplings~\eqref{flavorcoupl} are explicitly given by,
\begin{align}
    g_{11} & = I_{ud}  \label{g11}  \\[0.3em]
    g_{00} & =  \tfrac{1}{4}\, (I_{uu} + I_{dd} + I_{ss}+ I_{cc} ) \, ,\\[0.3em]
    g_{33} & =  \tfrac{1}{2}\, (I_{uu} + I_{dd} ) \, , \\[0.3em]
    g_{44} & =  I_{us}   \, , \\[0.3em]
    g_{66} & =  I_{ds}  \, , \\[0.3em]    
    g_{88} & =  \tfrac{1}{6}\, ( I_{uu} + I_{dd} + 4 I_{ss} )  \, , \\[0.3em]
    g_{99}  & =   I_{uc}   \, , \\[0.3em]
    g_{1111} & =   I_{dc}   \, , \\[0.3em]
    g_{1313} & =  I_{sc}   \, , \\[0.3em]
    g_{1515} & =  \tfrac{1}{12}\, ( I_{uu} + I_{dd} + I_{ss}+ 9 I_{cc} ) \, , \label{g1515}
\end{align}
where we employ the following standard correspondence of flavor eigenstates with the coupling constants~\cite{ParticleDataGroup:2024cfk}:
\begin{eqnarray}
    (\pi^0) \, ,    (\pi^+, \pi^-)  &\  \Longleftrightarrow  \   & (g_{33})  \, ,    (g_{11},g_{22}) \, , \nonumber  \\
    (K^\pm) \, ,   (K^0,\bar{K}^0)  & \  \Longleftrightarrow \ & (g_{44}, g_{55}) \, ,   (g_{66}, g_{77}) \, , \nonumber \\
    (D^\pm) \,  ,  (D^0,\bar{D}^0)  & \  \Longleftrightarrow  \ &  (g_{1111}, g_{1212})  \, ,  (g_{99}, g_{1010}) \, , \nonumber  \\
    (D_s)  & \  \Longleftrightarrow\ & (g_{1313}, g_{1414}) \, ,  \nonumber \\
    (\eta_c)  & \ \Longleftrightarrow \ & (g_{1515} ) \, .
\end{eqnarray}
We remark that we do not consider $\eta_c-\eta-\eta'$ mixing, as the inclusion of light quarks was found to contribute a mass difference of merely  
$\Delta m_{\eta_c} \approx 11$\;MeV~\cite{Bali:2011rd}. Therefore, we assume that the $\eta_c$  does not correspond to an $ij = 15$ state in solving  
the BSE~\eqref{eq:BSEps} despite using that representation in the coupling. Nonetheless, allowing for mixing may increase the value of $m_{\eta_c}$  
which is underestimated in this approach; see Tab.~\ref{table1}. We remark that the inclusion of $\eta$ mesons has been investigated 
with CI models in  Refs.~\cite{Zamora:2023fgl,Xu:2024frc}.

The couplings $g_{ij}$ enter the BSE via the integral kernel~\eqref{eq:Kcontact}, where they replace the charge-independent couplings, 
$1/(m_{g,f} m_{g,e})$. More precisely, we define the coupling in the BSE of a charged pion as, 
\begin{equation}
   g_{\pi^\pm}  = g_0 + g_{11}  \, 
\label{gpi-couple}   
\end{equation}
and normalize the neutral pion coupling, 
\begin{equation}
  g_{\pi^0}  =  \frac{g_0( g_0 + g_{33} ) }{  g_{\pi^\pm} } \, ,
\end{equation}
such that both pion states are related. Clearly, for vanishing polarization, 
or in the case of degenerate quark masses, $g_{\pi^\pm} =  g_{\pi^0} \to g_0$,
and in general $g_{ii} \ll g_0$. 

Similarly, the kaon couplings are normalized with the charged pion coupling,
\begin{equation}
     g_{K^\pm}   =  \frac{g_0( g_0 + g_{44} ) }{ g_{\pi^\pm} }  \, ,   \ \     g_{K^0}  = \frac{g_0( g_0 + g_{66} ) }{  g_{\pi^\pm} }  \, ,  
\end{equation}
while we relate the coupling of the neutral to that of the charged $D$ mesons by, 
\begin{equation}     
     g_{D^\pm}   =  g_0 + g_{1111}   \, ,   \ \     g_{D^0}  = \frac{g_0( g_0 + g_{99} ) }{ g_{D^\pm}}  \, ,  
\end{equation}
and the $D_s$ meson coupling is given by,
\begin{equation}
    g_{D_s} = \frac{g_0( g_0 + g_{1313} ) }{ g_{D^\pm}} \, . 
\end{equation}
The coupling that enters the BSE of the $\eta_c$ is defined as, 
\begin{equation}
    g_{\eta_c} = \frac{g_0( g_0 + g_{1515})}{ g_{\pi^\pm}} \, . 
 \label{getac-couple}   
\end{equation}

These neutral-meson couplings can be related to the couplings in the gap equation~\eqref{CI-DSE} in another representation:
\begin{equation}
   g_{ij} ( \bar{\psi} \lambda_i \psi ) (\bar{\psi} \lambda_j \psi )  = 2\, \hat g_{f_1\! f_2}  (\bar{\psi} \psi )_{f_1}  (\bar{\psi} \psi )_{f_2} \, , 
\end{equation}
where $f_i = u,d,s,c$. Neglecting the mixing interactions, the flavor-singlet quark couplings,  $g_i \equiv 1/m_{g,i}^2$, that enter the gap equation are defined as:
\begin{align} 
  g_u \equiv \, \hat g_{uu} &= \tfrac{1}{2}\,  g_{00} + g_{33}  + \tfrac{1}{3} \, g_{88}   + \tfrac{1}{6} \, g_{1515} \, , 
\label{Guu}  
\\[0.3em]
  g_d \equiv \, \hat g_{dd} &= \tfrac{1}{2} \, g_{00} + g_{33}  + \tfrac{1}{3} \, g_{88}   + \tfrac{1}{6} \, g_{1515} \, , 
\label{Gdd}
\\[0.3em]
  g_s \equiv \, \hat g_{ss} &= \tfrac{1}{2} \, g_{00}  + \tfrac{4}{3} \, g_{88}  + \tfrac{1}{6} \, g_{1515} \, ,
 \label{Gss}
\\[0.3em]
  g_c \equiv \,  \hat g_{cc} &= \tfrac{1}{2} \, g_{00}  + \tfrac{3}{2} \, g_{1515} \,   .
\label{Gcc}
\end{align}
With the above couplings, we solve the gap equation~\eqref{CI-DSE} and the BSE~\eqref{eq:BSEps} for different flavor states in the pseudoscalar channel. 
The present contact interaction model therefore involves, besides the four quark masses and the free coupling $g_0$, the infrared and ultraviolet regulators 
$\Lambda_{\mathrm{ir}}$ and $\Lambda_{\mathrm{uv}}$, respectively. 
\vspace*{-2mm}
 
%%%%%%%%%%%%%%%%%%%%%%%%%%%%%%%%%%%%%%%%%%%%%%%%%%%%%%%%%%%%%%%%%%%%%%%%%%%%%%%%%%%%%%%
\begin{table}[t!]
\renewcommand{\arraystretch}{1.3}
\setlength{\tabcolsep}{6pt}
\centering
\begin{tabular}{c|c|c|c||c|c|c}
\hline\hline
 & $m_M$ &$m^\mathrm{ref.}_M$   &  $\epsilon_m$[\%]    &   $f_M $  &  $f^\mathrm{ref.}_M $  &  $\epsilon_f$[\%]  \\ \hline
 $\pi^\pm $  & 139.3  &  139.6  & 0.2 &  129.5 &  130.2  & 0.5  \\ 
 $\pi^0$      & 139.0  & 135.0  & 3.0 &  $-$ & $-$  & $-$  \\ 
 $K^\pm$    & 495.9  &  493.7  & 0.4 & $  $ 167.1 &  155.7  &  7.3 \\ 
 $K^0$      & 498.2  & 497.6   & 0.1 & $-$  &  $-$ &  $-$  \\ 
 $D^\pm$    & 1889.9 & 1869.5   & 1.1 &  215.6  & 212.0 & 1.7   \\ 
 $D^0$      & 1889.4  & 1864.8  & 1.3 &  $-$  &  $-$ & $-$   \\ 
 $D_s$      & 1941.8 & 1968.4  & 1.4 &   246.4   & 249.9 & 1.4   \\ 
 $\eta_c$   & 2798.0  & 2984.1  & 6.2 &  293.6  &  385.0 & 23.7   \\ 
\hline \hline
\end{tabular}
\caption{Masses and weak decay constants [in MeV] of pseudoscalar mesons. The reference values are worldwide averages of the Particle Data Group 
             [PDG]~\cite{ParticleDataGroup:2024cfk} except for the weak decay constant of the $\eta_c$~\cite{Davies:2010ip}. The relative deviations 
             are given by $\epsilon_v  = | v^\textrm{ref.} - v^\textrm{th.} | / v^\textrm{ref.}$\!.}
\label{table1}
\end{table}

%%%%%%%%%%%%%%%%%%%%%%%%%%%%%%%%%%%%%%%%%%%%%%%%%%%%%%%%%%%%%%%%%%%%%%%%%%%%%%%%%%%%%%%%%%

\begin{table}[t!]
\centering
\begin{tabular}{ C{1.1cm} |  C{1.4 cm}| C{1.4cm} | C{1.4cm} | C{1.4cm}  }
\hline \hline
 $f$ &   $u$  &  $d$  &  $s$   & $c $  \\
\hline
 %$m_f$  &  0.0010 & 0.0027 & 0.052 &  1.05  \\
 %$M_f$  &  0.20 & 0.22 & 0.44  & 1.61 \\
 % 
  $m_f$  &  1.3 & 2.2 & 52.0 &  970.0  \\
 $M_f$  &  244.9 & 254.0 & 480.8  & 1555.5 \\
\hline \hline
\end{tabular}
\caption{Current and constituent quark masses [in MeV] obtained with $g_0 = 0.195042$\;GeV$^{-2}$, $\Lambda_{\mathrm{ir}}= 0.240$\;GeV and 
              $\Lambda_{\mathrm{uv}}= 2.325$\;GeV.  }
\label{table2}                 
\end{table}
%%%%%%%%%%%%%%%%%%%%%%%%%%%%%%%%%%%%%%%%%%%%%%%%%%%%%%%%%%%%%%%%%%%%%%%%%%%%%%%%%%%%%%%%%%

%%%%%%%%%%%%%%%%%%%%%%%%%%%%%%%%%%%%%%%%%%%%%%%%%%%%%%%%%%%%%%%%%%%%%%%%%%%%%%%%%%%%%%%
%%%%%%%%%%%%%%%%%%%%%%%%%%%%%%%%%%%%%%%%%%%%%%%%%%%%%%%%%%%%%%%%%%%%%%%%%%%%%%%%%%%%%%%
 
\section{Results\label{results}}

Using the FSB couplings of the effective Lagrangian~\eqref{efflagrangianU4}, the gap equation~\eqref{CI-DSE} and the BSE~\eqref{eq:BSEps}, the mass spectrum  and the weak 
decay constants~\eqref{pion_cte_DS} of the pions, kaons, $D$ mesons and $\eta_c$ can be reproduced by adjusting the scale parameters $g_0$, $\Lambda_{\mathrm{ir}}$ and 
$\Lambda_{\mathrm{uv}}$ and the quark masses, $m_u$, $m_d$,  $m_s$ and $m_c$.

The light quark masses are initially chosen to be close to the Particle Data Group values~\cite{ParticleDataGroup:2024cfk} at 2\;GeV and we only use one pair 
of infrared and ultraviolet regulators, that is both the light and heavy sectors are treated on equal footing contrary to Ref.~\cite{Serna:2017nlr}. We initially fix 
$m_u$, $m_d$ and $\Lambda_{\mathrm{ir}}$, whereas $m_s$, $m_c$, $g_0$ and $\Lambda_{\mathrm{uv}}$ are obtained in a least-square fit to the worldwide 
averaged values~\cite{ParticleDataGroup:2024cfk} of pseudoscalar meson masses and weak decay constants listed in Tab.~\ref{table1}. The light-quark masses 
are then adjusted so as to produce a reasonable mass splitting of the pions due to strong interactions~\cite{Giusti:2017dmp}. The resulting masses and scale parameters 
are detailed in Tab.~\ref{table2}, while the corresponding BSAs and normalized flavored coupling constants for each meson are listed in Tabs.~\ref{tableBSA} and
 \ref{table:Gij}, respectively.

%%%%%%%%%%%%%%%%%%%%%%%%%%%%%%%%%%%%%%%%%%%%%%%%%%%%%%%%%%%%%%%%%%%%%%%%%%%%%%%%%%%%%%%%%%
\begin{table}[t!]
\centering
\setlength{\tabcolsep}{4.5pt}
\renewcommand{\arraystretch}{1.5}
\begin{tabular}{c|c|c|c||c|c|c|c}
\hline \hline
 $E_{\pi^\pm}$ & $F_{\pi^\pm}$  & $E_{\pi^0}$ & $F_{\pi^0}$ & $E_{K^\pm}$ & $F_{K^\pm}$  & $E_{K^0}$ & $F_{K^0}$ \\
\hline
  1.909  &  0.041 & 1.909 & 0.042 & 1.983 & 0.070 & 1.988 & 0.072 \\
\hline \hline
 $E_{D^\pm}$ & $F_{D^\pm}$  & $E_{D^0}$ & $F_{D^0}$ & $E_{D_s}$ & $F_{D_s}$  & $E_{\eta_c}$ & $F_{\eta_c}$ \\
 \hline 
2.048 & 0.122  & 2.036 & 0.117 & 2.342 & 0.233 & 3.033 & 0.585   \\
\hline\hline  
\end{tabular}
\caption{The Bethe-Salpeter amplitudes~\eqref{contactBSA} of the $\pi^\pm$, $\pi^0$, $K^\pm$, $K^0$, $D^\pm$, $D^0$, $D_s$ and $\eta_c$ that yield the meson masses and weak decay constants listed in Tab.~\ref{table1}.}
\label{tableBSA}     
\vspace*{-3mm}
\end{table}
%%%%%%%%%%%%%%%%%%%%%%%%%%%%%%%%%%%%%%%%%%%%%%%%%%%%%%%%%%%%%%%%%%%%%%%%%%%%%%%%%%%%%%%%%%

%%%%%%%%%%%%%%%%%%%%%%%%%%%%%%%%%%%%%%%%%%%%%%%%%%%%%%%%%%%%%%%%%%%%%%%%%%%%%%%%%%%%%%%%%%
\begin{table}[t!]
\centering
\setlength{\tabcolsep}{8pt}
\renewcommand{\arraystretch}{1.5}
\begin{tabular}{c|c|c|c}
\hline \hline
  $g_{\pi^\pm}$ & $g_{\pi^0}$ & $g_{K^\pm}$ & $g_{K^0}$ \\
\hline
  0.195042 & 0.195042 & 0.195035 & 0.195036  \\
\hline \hline 
  $g_{D^\pm}$ & $g_{D^0}$  & $g_{D_s}$ & $g_{\eta_c}$ \\
\hline   
  0.195042 & 0.195038 & 0.195272 &  0.194629  \\
\hline \hline 
 $g_u$ & $g_d$  & $g_s$ & $g_c$ \\
 \hline 
 0.194985 & 0.194985 & 0.195008 & 0.194702 \\
\hline \hline
\end{tabular}
\caption{The charge-dependent couplings defined in Eqs.~\eqref{gpi-couple} to \eqref{getac-couple} with $g_0 = 0.195042$\;GeV$^{-2}$ and the gap-equation couplings 
for the four flavors defined in Eqs.~\eqref{Guu} to~\eqref{Gcc} (in GeV$^{-2}$).} 
\label{table:Gij}   
\end{table}
%%%%%%%%%%%%%%%%%%%%%%%%%%%%%%%%%%%%%%%%%%%%%%%%%%%%%%%%%%%%%%%%%%%%%%%%%%%%%%%%%%%%%%%%%%

With the exception of the weak decay constant of the $\eta_c$, the CI model augmented by flavor-dependent couplings reproduces rather well the experimental masses and 
weak decay constants. The light-quark mass ratio, $m_d/m_u = 1.7$, we choose yields $\Delta m_\pi = m_{\pi^+} - m_{\pi^0} \approx 0.3$\;MeV, which is somewhat above 
what lattice QCD~\cite{Giusti:2017dmp} and chiral perturbation theory~\cite{Gasser:1982ap,Donoghue:1989sj} predict for the strong contributions to the masses. 
We remind that electromagnetic corrections have been neglected in this work and that this mass splitting is due to the explicit isospin breaking  \emph{and} to the polarization 
corrections, Eqs.~\eqref{g11} to \eqref{g1515}.  The contributions of the latter are more important for $m_d/m_u \gtrsim 2$. However, even in the present case, if merely $g_0$ 
enters the gap and bound-state equations as a universal coupling strength, the mass splitting is $\Delta m_\pi \approx 0.27$\;MeV and the $D^\pm$ and $D^0$ mesons are 
mass degenerate.  

The kaon mass difference is $\Delta m_K = m_{K^0} - m_{K^\pm} = 2.3$\;MeV and the ratio $(m_{K^0}^2 - m_{K^\pm}^2)/(m_d-m_u) = 2.54$\;GeV is in excellent agreement 
with  the value at the physical point in Ref.~\cite{Giusti:2017dmp}. Similarly, in the case of the $D$ mesons, $m_{D^\pm}^2 - m_{D^0}^2 \approx 0.002$\;GeV$^2$, while in  
Ref.~\cite{Giusti:2017dmp} this value is about $0.012$\;GeV$^2$. We note that these mass splittings cannot be obtained with an earlier version of the CI model~\cite{Serna:2017nlr} 
which also did not consider the $\eta_c$.

As noted earlier, in the CI model the weak decay constant of the $\eta_c$, for which no experimental data exists, is much smaller than a prediction of lattice QCD~\cite{Davies:2010ip} 
and its mass is about 180\;MeV too low. An increase of $m_c$ leads to an $\eta_c$ mass in better agreement with the experimental value, but also increases the masses of the 
$D$ mesons by about 60\;MeV. One may, of course, question whether the use of  a single coupling, $g_0 = 1/\surd{m_g} \approx 2.3$\;GeV, captures the large-scale difference 
between the light and charm quarks, in particular in the case of charmonia. Further investigations should include the $J/\psi$, for instance, as well as the $D^*$ and $D_s^*$ mesons.

%%%%%%%%%%%%%%%%%%%%%%%%%%%%%%%%%%%%%%%%%%%%%%%%%%%%%%%%%%%%%%%%%%%%%%%%%%%%%%%%%%%%%%%
%%%%%%%%%%%%%%%%%%%%%%%%%%%%%%%%%%%%%%%%%%%%%%%%%%%%%%%%%%%%%%%%%%%%%%%%%%%%%%%%%%%%%%%

\section{Final Remarks\label{lastsec}}

The CI model has been successfully applied to compute the mass spectrum of light and heavy pseudoscalar mesons and their associated decay and coupling constants. 
The model can also be used to predict elastic and transition form factors at low momentum transfer. This is commonly done in the isospin limit $m_u= m_d$, where the 
charge states of the hadrons are irrelevant and the couplings are charge independent. The aim of this work is twofold: \emph{i}) reduce the number of the model's parameters 
and \emph{ii}) study the effect of one-loop FSB polarization corrections to a universal coupling derived with the background field method.  

Our findings reveal that the number of free parameters of the CI model in Ref.~\cite{Serna:2017nlr} can be limited to the quark masses, a single pair of infrared and ultraviolet 
cutoffs, and an overall mass scale. They suffice to reproduce rather well the experimental values of the pion, kaon, and $D$ mesons. In addition, we included the $\eta_c$ 
not considered previously, though a consistent fit of the lighter meson and the charmonium is not achievable with this restricted set of parameters. Taking into consideration 
a previous study~\cite{Serna:2017nlr}, this suggests that the treatment of heavier quarkonia and $B$ mesons requires an additional mass scale in the CI model.

An additional improvement over Ref.~\cite{Serna:2017nlr} is the introduction of FSB couplings, whose flavor dependence stems from polarization corrections and which 
replace couplings that are individually fitted to data. Studies of the quark-gluon vertex unambiguously demonstrate the flavor content of the vertex dressing and of its suppression 
in the case of heavy flavors~\cite{Albino:2021rvj,El-Bennich:2022obe,Lessa:2022wqc}. These strong-interaction effects must be taken into account in effective couplings. 
Their definition in the BSE for a given isospin and flavor, motivated by the strong corrections obtained with a mean-field approach, relies solely on a single mass scale to be fitted, 
and with the typical explicit isospin breaking, $m_d \approx 2 m_u$, we find a mass difference of 0.3\;MeV between the charged and neutral pions. This mass splitting is
consistent with that found in lattice QCD calculations considering only strong interaction effects. The differences between charged and neutral kaons and $D$ mesons, 
however, appear to be underestimated in this approach and higher-order corrections can be important. 

Given the usefulness of the CI model despite its simplicity, as explained in Section~\ref{sec:intro}, it is important to reduce its number of parameters to a necessary minimum 
guided by flavor symmetry and its breaking patterns. The present work leads to a more consistent calculation of the pseudoscalar meson spectrum, as the strong FSB 
couplings are not separately adjusted in the light and heavy flavor sectors. Instead, they are related to a single gluon-mass scale via polarization integrals introduced in Refs.~\cite{Braghin:2021,Braghin:2021hmr}. As mentioned before, the present CI model ought to be applied to the vector-meson channel and beyond the charm sector, 
which implies the use of flavored SU(5) couplings to include the $B$ mesons and the $\Upsilon$. 

%%%%%%%%%%%%%%%%%%%%%%%%%%%%%%%%%%%%%%%%%%%%%%%%%%%%%%%%%%%%%%%%%%%%%%%%%%%%%%%%%%%%%%%

\acknowledgements

Financial support by CNPq is acknowledged, grant nos.~312750/2021-8 and 407162/2023-2 (F.~L.~B.),  grant no.~2023/00195-8 (B.~E. and F.~E.~S.) and by FAPESP, 
grant no.~2023/00195-8 (B.~E.). 

%%%%%%%%%%%%%%%%%%%%%%%%%%%%%%%%%%%%%%%%%%%%%%%%%%%%%%%%%%%%%%%%%%%%%%%%%%%%%%%%%%%%%%%

\end{document}